\documentclass[aps,pra,twocolumn,superscriptaddress]{revtex4-1}
\usepackage{graphicx,amsmath,amssymb}
\usepackage{subfigure}
\usepackage{float}
\usepackage{multirow}

\begin{document}

\title{Quantum phase transitions and critical behaviors in the two-mode three-level quantum Rabi model}

\author{Yan Zhang}
\affiliation{Beijing Computational Science Research Center, Beijing 100193, China}
\affiliation{College of Physics and Electronic Engineering, Northwest Normal University, Lanzhou 730070, China}

\author{Bin-Bin Mao}
\affiliation{School of Physical Science and Technology $\&$ Key Laboratory for Magnetism and Magnetic Materials of the Ministry of Education, Lanzhou University, Lanzhou 730000, China}
\affiliation{Department of Physics, The University of Hong Kong, Hong Kong, China}

\author{Dazhi Xu}
\affiliation{Center for Quantum Technology Research and School of Physics, Beijing Institute of Technology, Beijing 100081, China}

\author{Yu-Yu Zhang}
\affiliation{Department of Physics, Chongqing University, Chongqing 401330, China}

\author{Wen-Long You}
\email{youwenlong@gmail.com}
\affiliation{College of Science, Nanjing University of Aeronautics and Astronautics, Nanjing 211106, China}
\affiliation{School of Physical Science and Technology, Soochow University, Suzhou, Jiangsu 215006, China}

\author{Maoxin Liu}
\email{liumaoxin@bupt.edu.cn}
\affiliation{State Key Laboratory of Information Photonics and Optical Communications $\&$ School of Science, Beijing University of Posts and Telecommunications, Beijing 100876, China}
\affiliation{Beijing Computational Science Research Center, Beijing 100193, China}

\author{Hong-Gang Luo}
\email{luohg@lzu.edu.cn}
\affiliation{School of Physical Science and Technology $\&$ Key Laboratory for Magnetism and Magnetic Materials of the Ministry of Education, Lanzhou University, Lanzhou 730000, China}
\affiliation{Beijing Computational Science Research Center, Beijing 100193, China}

\begin{abstract}
We explore an extended quantum Rabi model describing the interaction between a two-mode bosonic field and a three-level atom. Quantum phase transitions of this few degree of freedom model is found when the ratio $\eta$ of the atom energy scale to the bosonic field frequency approaches infinity. An analytical solution is provided when the two lowest-energy levels are degenerate. According to it, we recognize that the phase diagram of the model consists of three regions: one normal phase and two superradiant phases. The quantum phase transitions between the normal phase and the two superradiant phases are of second order relating to the spontaneous breaking of the discrete $Z_{2}$ symmetry. On the other hand, the quantum phase transition between the two different superradiant phases is discontinuous with a phase boundary line relating to the continuous $U(1)$ symmetry. For a large enough but finite $\eta$, the scaling function and critical exponents are derived analytically and verified numerically, from which the universality class of the model is identified.
\end{abstract}

\maketitle

\section{Introduction}
The quantum Rabi model describes the interaction between a photon field and a two-level system \cite{Rabi1936,Rabi1937}, which is the simplest model for studying the light-matter interaction and plays a significant role in quantum optics \cite{Jaynes1963}, condensed-matter physics \cite{Holstein1959}, and quantum information \cite{Raimond2001}. With the rapid experimental progress in accessing the strong \cite{Blais2004,Wallraff2004}, the ultrastrong \cite{Niemczyk2010,Forn-Diaz2010,Chen2017}, and the deep strong coupling regimes \cite{Forn-Diaz2017,Yoshihara2017}, the quantum Rabi model has received much attention since the rotating-wave approximation fails \cite{Irish2007,Casanova2010,Gan2010,Braak2011,Larson2012,Yu2012,Chen2012,Ashhab2013,Lee2013,Liberato2014,Xie2014,Liu2015,Ying2015,Cong2017,Zhang2017,Wang2018,Mao2019,Kockum2019,Forn-Diaz2019}. Recently, phase transitions and critical phenomena have been surprisingly found in the quantum Rabi model although only a single atom is involved \cite{Hwang2015}, which requires the infinite frequency ratio of the atom to the photon rather than the thermodynamic limit required by traditional phase transitions. Further study on the scaling behaviors of the Rabi and the Dicke models has revealed that these two models belong to the same universality class \cite{Liu2017}. These progresses bring a new insight for the quantum phase transition without the thermodynamic limit.

In parallel the interaction between a two-mode cavity field and a three-level system leads to many important phenomena, such as electromagnetically induced transparency \cite{Boller1991} and dark state \cite{Fleischhauer2000}, which are profitable in the precise control of coherent population trapping and transfer \cite{Bergmann1998}. The three-level system is also important in quantum information, referred as qutrit. Compared with the two-level scheme, the quantum key distribution based on qutrits is more resistant to attack \cite{BruB2002,Cerf2002}, and the quantum computation using qutrits shows a faster speed and a lower error rate \cite{Zhou2002,Sjoqvist2012}. A qutrit quantum computer with trapped ions has been proposed \cite{Klimov2003}. In addition, the three-level system is used to construct a quantum heat engine \cite{Geva1996,Xu2016}. To identify the possible quantum phases and quantum phase transitions involved in a two-mode three-level model is helpful in further understanding these light-matter interaction models and extending their applications.

The two-mode three-level interaction model in the thermodynamic limit has received much attention. Hayn \textit{et al.} studied quantum phase transitions by a generalized Holstein-Primakoff transformation and revealed that it exhibits two superradiant quantum phase transitions, which can be both first and second order \cite{Hayn2011}. Cordero \textit{et al.} found that the polychromatic ground-state phase diagram can be divided into monochromatic regions by a variational analysis \cite{Cordero2015}. Here, we report an analytical calculation of the ground-state phase diagram, scaling function, and critical exponents for the two-mode three-level quantum Rabi model by taking a single $\Lambda$-type three-level atom as a prototype. The analytical results are further verified by an exact numerical diagonalization.

The paper is organized as follows. In Sec. II, an effective model is derived when the ratio between the atom frequency and the photon frequency approaches infinity. In Sec. III, a ground-state phase diagram is extracted analytically. In Sec. IV, the mean photon number in the ground state is analytically derived. In Sec. V, the scaling function and critical exponents are analytically derived for finite frequency ratios, and also the numerical diagonalization is used to verify the analytical results. Finally, a brief summary is presented in Sec. VI.

\section{Model Hamiltonian}
Our two-mode three-level quantum Rabi model describes the interaction between a two-mode quantized field and a single $\Lambda$-type three-level atom, which is given by ($\hbar=1$)
\begin{eqnarray}
\hat{H}=&&\varepsilon_{1}|1\rangle\langle1|+\varepsilon_{2}|2\rangle\langle2|+\varepsilon_{3}|3\rangle\langle3| +\omega_{1}\hat{a}_{1}^{\dagger}\hat{a}_{1}+\omega_{2}\hat{a}_{2}^{\dagger}\hat{a}_{2} \nonumber \\
&&+g_{1}\hat{A}_{1}(\hat{a}_{1}^{\dagger}+\hat{a}_{1})+g_{2}\hat{A}_{2}(\hat{a}_{2}^{\dagger}+\hat{a}_{2}),
\end{eqnarray}
where $\hat{A}_{1}=|1\rangle\langle3|+|3\rangle\langle1|$, $\hat{A}_{2}=|2\rangle\langle3|+|3\rangle\langle2|$, $\varepsilon_{i}$ is the energy eigenvalue of state $|i\rangle$ ($i=1, 2, 3$), $\hat{a}_{1}^{\dagger}$ and $\hat{a}_{1}$ ($\hat{a}_{2}^{\dagger}$ and $\hat{a}_{2}$) are the creation and annihilation operators of the photon mode 1 (mode 2), $\omega_{1}$ ($\omega_{2}$) is the frequency of the photon mode 1 (mode 2), $g_{1}$ ($g_{2}$) is the coupling strength between the transition $|1\rangle\leftrightarrow|3\rangle$ ($|2\rangle\leftrightarrow|3\rangle$) and the photon mode 1 (mode 2). Note that the transition between state $|1\rangle$ and state $|2\rangle$ in the $\Lambda$-type configuration is forbidden.

For convenience, we define several variables to make the Hamiltonian dimensionless: $\varDelta=\varepsilon_{3}-\varepsilon_{1}$, $\delta=(\varepsilon_{2}-\varepsilon_{1})/\varDelta$, $\alpha=\omega_{2}/\omega_{1}$, $\beta=g_{2}/g_{1}$, $R=2g_{1}/\sqrt{\omega_{1}\varDelta}$. Also, we set $\varepsilon_{1}=0$ and $\eta=\varDelta/\omega_{1}$, and then the dimensionless Hamiltonian rescaled by $\varDelta$ becomes
\begin{eqnarray}
\hat{H}'=&&\delta|2\rangle\langle2|+|3\rangle\langle3|+\frac{1}{\eta}(\hat{a}_{1}^{\dagger}\hat{a}_{1}+\alpha \hat{a}_{2}^{\dagger}\hat{a}_{2}) \nonumber \\
&&+\frac{1}{2\eta^{1/2}}R[\hat{A}_{1}(\hat{a}_{1}^{\dagger}+\hat{a}_{1})+\beta \hat{A}_{2}(\hat{a}_{2}^{\dagger}+\hat{a}_{2})].
\end{eqnarray}
We rewrite the Hamiltonian in terms of the position and momentum operators ($\hat{x}_i$ and $\hat{p}_{x_i}$, $i=1, 2$) via $\hat{a}_i=(\hat{x}_i+i\hat{p}_{x_i})/\sqrt{2}$ and $\hat{a}^{\dagger}_i=(\hat{x}_i-i\hat{p}_{x_i})/\sqrt{2}$. Further rescaling the position and momentum operators by $\hat{y}_i=\eta^{-1/2}\hat{x}_i$, the Hamiltonian becomes
\begin{eqnarray}\label{Hpm}
\hat{H}'=&&\delta|2\rangle\langle2|+|3\rangle\langle3|+\frac{1}{2}(\hat{y}_{1}^{2}+\alpha \hat{y}_{2}^{2})+\frac{1}{2\eta^{2}}(\hat{p}_{y_{1}}^{2}+\alpha \hat{p}_{y_{2}}^{2}) \nonumber \\
&&+\frac{\sqrt{2}}{2}R(\hat{A}_{1}\hat{y}_{1}+\beta \hat{A}_{2}\hat{y}_{2}).
\end{eqnarray}
Supposing that $\hat{p}_{y_{1}}$ and $\hat{p}_{y_{2}}$ are finite, the contributions from the momentum terms disappear in the limit $\eta\rightarrow\infty$, and then the effective Hamiltonian reduces to
\begin{eqnarray}\label{Hlimit}
\hat{H}_{\mathrm{eff}}=&&\delta|2\rangle\langle2|+|3\rangle\langle3|+\frac{1}{2}(y_{1}^{2}+\alpha y_{2}^{2}) \nonumber \\
&&+\frac{\sqrt{2}}{2}R(\hat{A}_{1}y_{1}+\beta \hat{A}_{2}y_{2}).
\end{eqnarray}
Because of the absence of momentum operators, $\hat{y}_{1}$ and $\hat{y}_{2}$ can be replaced
with the eigenvalues $y_{1}$ and $y_{2}$ in Eq.~\eqref{Hlimit}. In consequence, the ground-state phase diagram and the mean photon number are obtained.

\section{Phase diagram}

\begin{figure}
  \centering
  \includegraphics[width=6cm]{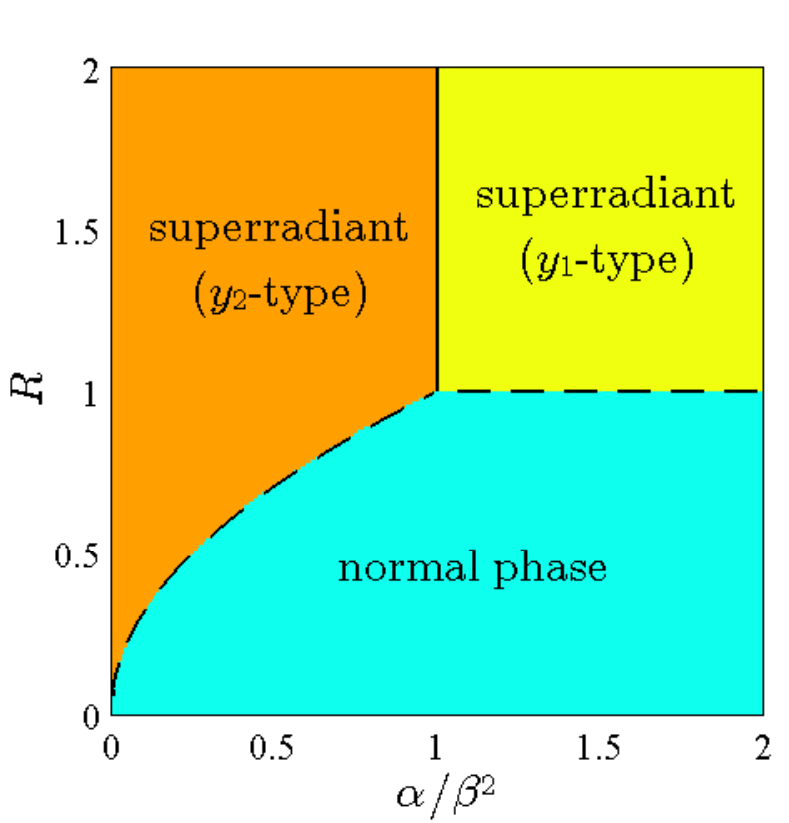}\\
  \caption{Ground-state phase diagram as a function of the parameter ratio $\alpha/\beta^{2}$ and the coupling strength $R$. The lower part is the normal phase. The higher part is the superradiant phase with two regimes of $\alpha<\beta^{2}$ and $\alpha>\beta^{2}$. The former is only contributed to by the coupling of mode 2 and transition $|2\rangle\leftrightarrow|3\rangle$, labelled as $y_{2}$-type. The latter is only from the coupling of mode 1 and transition $|1\rangle\leftrightarrow|3\rangle$, labelled as $y_{1}$-type. Two distinct regions are separated by a border of $\alpha=\beta^{2}$. The solid and dashed lines represent the first-order and second-order phase transitions, respectively.}
  \label{PhaseDiagram}
\end{figure}

The analytical diagonalization of the effective Hamiltonian \eqref{Hlimit} is equivalent to solving a third-order algebraic equation. An explicit form of the ground-state energy can be readily obtained for a special case that the state $|1\rangle$ and the state $|2\rangle$ are degenerate (i.e., $\delta=0$),
\begin{equation}\label{Eg}
E(y_{1},y_{2})=\frac{1}{2}(y_{1}^{2}+\alpha y_{2}^{2})+\frac{1}{2}[1-\sqrt{1+2R^{2}(y_{1}^{2}+\beta^{2}y_{2}^{2})}].
\end{equation}
Through the first-order and the second-order derivatives ($\frac{\partial E}{\partial y_{1}}$, $\frac{\partial E}{\partial y_{2}}$; $\frac{\partial^{2} E}{\partial y_{1}^{2}}$, $\frac{\partial^{2} E}{\partial y_{2}^{2}}$), the ground-state energy $E_{0}$ is further determined. The detailed derivation of the ground-state energy is given in Appendix A. The obtained ground-state phase diagram is depicted in Fig.~\ref{PhaseDiagram}. One can see that the phase diagram shows several distinct regimes.

\begin{figure}
\centering
\subfigure[\ $\frac{\partial^{2} E_{0}}{\partial R^{2}}$]{
\begin{minipage}{4cm}
\centering
\includegraphics[width=4.2cm]{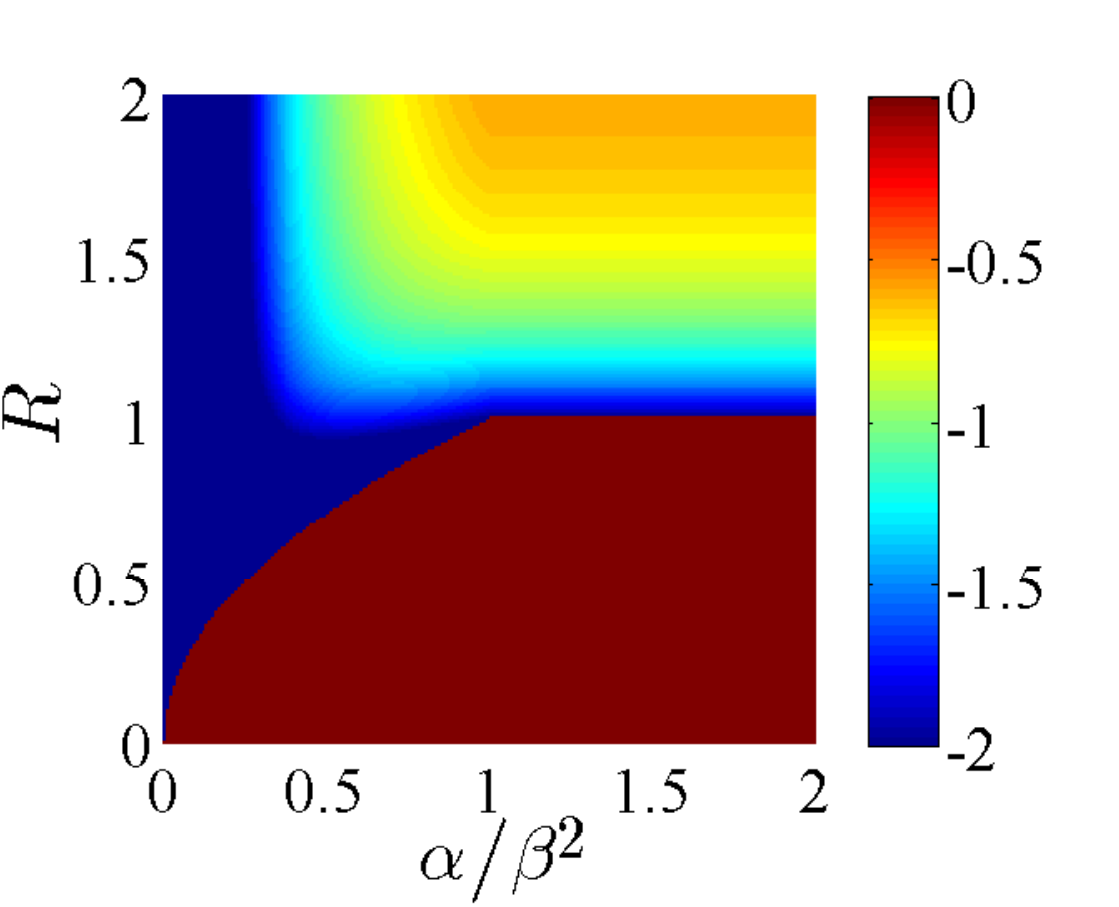}
\end{minipage}
}
\subfigure[\ $\frac{\partial E_{0}}{\partial \gamma}$]{
\begin{minipage}{4cm}
\centering
\includegraphics[width=4.2cm]{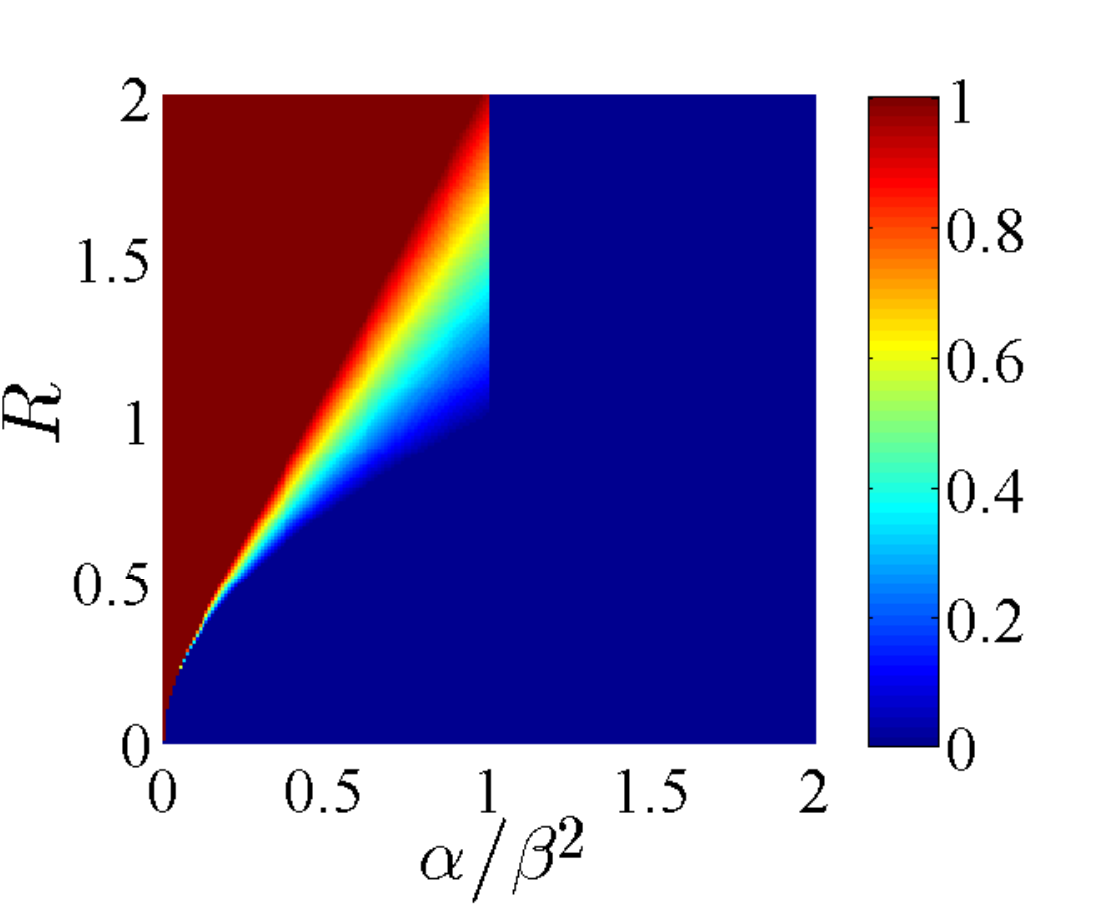}
\end{minipage}
}
\caption{(a) $\frac{\partial^{2} E_{0}}{\partial R^{2}}$, the second-order derivative of the ground-state energy with respect to the coupling strength $R$. It reflects a second-order normal-superradiant phase transition at the critical points. (b) $\frac{\partial E_{0}}{\partial \gamma}$, the first-order derivative of the ground-state energy with respect to the parameter ratio $\gamma=\alpha/\beta^{2}$, reflecting a first-order transition between the two superradiant phases of $\alpha<\beta^{2}$ and $\alpha>\beta^{2}$.}
\label{PhaseDiagram-Order}
\end{figure}

In the regime of $\alpha<\beta^{2}$, the critical point of phase transition is
\begin{equation}
R_{<,c}=\sqrt{\frac{\alpha}{\beta^{2}}}.
\end{equation}
When $R$ is less than or equal to $R_{<,c}$, $y_{1}=0$ and $y_{2}=0$, and the ground state stays in a normal phase with $E_{0}=0$. When $R$ is larger than $R_{<,c}$, the normal phase becomes unstable and bifurcates into two degenerate stable solutions
\begin{equation}
y_{1}=0, \quad y_{2,\pm}=\pm\sqrt{\frac{\beta^{4}R^{4}-\alpha^{2}}{2\alpha^{2}\beta^{2}R^{2}}},
\end{equation}
with the ground-state energy
\begin{equation}
E_{0}=-\frac{1}{4}(\frac{\alpha}{\beta^{2}R^{2}}+\frac{\beta^{2}R^{2}}{\alpha})+\frac{1}{2}.
\end{equation}
It implies that the ground state enters the so-called superradiant phase. While $E_{0}$ and $\frac{\partial E_{0}}{\partial R}$ are continuous, $\frac{\partial^{2} E_{0}}{\partial R^{2}}$ is discontinuous at $R=R_{c}$ as shown in Fig.~\ref{PhaseDiagram-Order}(a), which reveals that the phase transition from the normal phase to the superradiant phase is of second order.

In the regime of $\alpha>\beta^{2}$, the critical point of phase transition is
\begin{equation}
R_{>,c}=1.
\end{equation}
When $R$ is less than or equal to $R_{>,c}$, the ground state corresponds to $E_{0}=0$ with $y_{1}=0$ and $y_{2}=0$, which is the normal phase. When $R$ is larger than $R_{>,c}$, the ground state bifurcates into two degenerate stable solutions
\begin{equation}
y_{1,\pm}=\pm\sqrt{\frac{R^{4}-1}{2R^{2}}}, \quad y_{2}=0,
\end{equation}
with the ground-state energy
\begin{equation}
E_{0}=-\frac{1}{4}(\frac{1}{R^{2}}+R^{2})+\frac{1}{2}.
\end{equation}
The ground-state energy is a constant in this case, independent of $\alpha$ and $\beta$. Likewise, the normal-superradiant quantum phase transition is also a second-order phase transition.

When $\alpha=\beta^{2}$, both $y_{1}$ and $y_{2}$ are nonzero in the ground state when $R$ exceeds the critical point $R_{=,c}=1$, which satisfy the relation:
\begin{equation}
y_{1}^{2}+\alpha y_{2}^{2}=\frac{R^{4}-1}{2R^{2}}.
\end{equation}
The ground-state energy is accordingly given by
\begin{equation}
E_{0}=-\frac{1}{4}(\frac{1}{R^{2}}+R^{2})+\frac{1}{2}.
\end{equation}
This quantum phase transition still has the second-order nature. However, across the boundary line ($\alpha=\beta^{2}$) between the two superradiant phases, the model undergoes a first-order quantum phase transition because the first-order partial derivative $\frac{\partial E_{0}}{\partial\gamma}$ ($\gamma=\alpha/\beta^{2}$) is discontinuous, as shown in Fig.~\ref{PhaseDiagram-Order}(b).
Here, the two-mode three-level quantum Rabi model presents two kinds of typical spontaneous symmetry breaking, which are separately characterized by the order parameters $y_{1}$ and $y_{2}$. One can see that the normal phase possesses a discrete $Z_{2}$ symmetry, but the ground-state energy functionals either $E(y_{2})$ or $E(y_{1})$ shows a double-well structure when the coupling strength exceeds the critical points, which implies a breaking of the $Z_{2}$ symmetry. In the case of $\alpha=\beta^{2}$, the ground-state energy functional $E(y_{1},y_{2})$ presents a continuous $U(1)$ symmetry. When the coupling strength is above the critical point, both $y_{1}$ and $y_{2}$ become nonzero, which reflects a breaking of the continuous symmetry. The $U(1)$ symmetry was also found in the two-level system \cite{Baksic2014}, in which the light-atom interaction, however, requires the consideration of both the electric and the magnetic components for the electromagnetic field. Such continuous symmetry breaking is related to the Nambu-Goldstone mode \cite{Fan2014}.

\section{Mean photon number}
The photon number is a common observable in experiments and usually used to characterize the states of the light-matter interaction systems. In the Dicke quantum phase transition experiments, an abrupt increase of the mean intracavity photon number marks the onset of the normal-superradiant phase transition \cite{Baumann2010,Baumann2011}. To calculate the mean photon number of the ground state, we rewrite the Hamiltonian in Eq.~\eqref{Hlimit} as (with $\delta=0$),
\begin{equation}
\hat{H}_{\mathrm{eff}}=\frac{1}{2}(y_{1}^{2}+\alpha y_{2}^{2})+\frac{1}{2}\hat{M},
\end{equation}
where
\begin{equation}
\hat{M}=\left(\begin{array}{ccc}
0 & 0 & \sqrt{2}Ry_{1}\\
0 & 0 & \sqrt{2}\beta Ry_{2}\\
\sqrt{2}Ry_{1} & \sqrt{2}\beta Ry_{2} & 2
\end{array}\right).
\end{equation}

When $\alpha<\beta^{2}$, the lowest eigenvalue of $\hat{M}$ is $1-\frac{\beta^{2}R^{2}}{\alpha}$ for $y_{1}=0$ and $y_{2,\pm}=\pm\sqrt{\frac{\beta^{4}R^{4}-\alpha^{2}}{2\alpha^{2}\beta^{2}R^{2}}}$, and the corresponding eigenstate (normalized) is
\begin{equation}\label{less-eigenstate}
\left(\begin{array}{c}
a_{2,\pm}\\
b_{2,\pm}\\
c_{2,\pm}
\end{array}\right)=\pm\left(\begin{array}{c}
0\\
-\sqrt{\frac{\beta^{2}R^{2}+\alpha}{2\beta^{2}R^{2}}}\\
\sqrt{\frac{\beta^{2}R^{2}-\alpha}{2\beta^{2}R^{2}}}
\end{array}\right).
\end{equation}
Thus the ground-state wavefunction of $\hat{H}_{\mathrm{eff}}$ can be expressed as
\begin{equation}
\psi_{2}=\frac{1}{\sqrt{2}}\left[\left(\begin{array}{c}
a_{2,+}\\
b_{2,+}\\
c_{2,+}
\end{array}\right)
+\left(\begin{array}{c}
a_{2,-}\\
b_{2,-}\\
c_{2,-}
\end{array}\right)\right].
\end{equation}
Based on the wavefunction $\psi_{2}$, the mean photon numbers of the two modes above the critical point are
\begin{equation}
\langle\hat{a}_{1}^{\dagger}\hat{a}_{1}\rangle=\frac{1}{2}\eta\langle\psi_{2}|\hat{y}_{1}^{2}|\psi_{2}\rangle=0
\end{equation}
and
\begin{equation}
\langle\hat{a}_{2}^{\dagger}\hat{a}_{2}\rangle=\frac{1}{2}\eta\langle\psi_{2}|\hat{y}_{2}^{2}|\psi_{2}\rangle=\frac{1}{2}\eta(y_{2,+}^{2}+y_{2,-}^{2})=\eta\frac{\beta^{4}R^{4}-\alpha^{2}}{4\alpha^{2}\beta^{2}R^{2}}.
\end{equation}

When $\alpha>\beta^{2}$, the lowest eigenvalue of $\hat{M}$ is $1-R^{2}$ for $y_{1,\pm}=\pm\sqrt{\frac{R^{4}-1}{2R^{2}}}$ and $y_{2}=0$, and the corresponding eigenstate (normalized) is
\begin{equation}\label{larger-eigenstate}
\left(\begin{array}{c}
a_{1,\pm}\\
b_{1,\pm}\\
c_{1,\pm}
\end{array}\right)=\pm\left(\begin{array}{c}
-\sqrt{\frac{R^{2}+1}{2R^{2}}}\\
0\\
\sqrt{\frac{R^{2}-1}{2R^{2}}}
\end{array}\right).
\end{equation}
Hence, the ground-state wavefunction of $\hat{H}_{\mathrm{eff}}$ can be expressed as
\begin{equation}
\psi_{1}=\frac{1}{\sqrt{2}}\left[\left(\begin{array}{c}
a_{1,+}\\
b_{1,+}\\
c_{1,+}
\end{array}\right)
+\left(\begin{array}{c}
a_{1,-}\\
b_{1,-}\\
c_{1,-}
\end{array}\right)\right].
\end{equation}
Then, the mean photon numbers of the two modes above the critical point are
\begin{equation}
\langle\hat{a}_{1}^{\dagger}\hat{a}_{1}\rangle=\frac{1}{2}\eta\langle\psi_{1}|\hat{y}_{1}^{2}|\psi_{1}\rangle=\frac{1}{2}\eta(y_{1,+}^{2}+y_{1,-}^{2})=\eta\frac{R^{4}-1}{4R^{2}}
\end{equation}
and
\begin{equation}
\langle\hat{a}_{2}^{\dagger}\hat{a}_{2}\rangle=\frac{1}{2}\eta\langle\psi_{1}|\hat{y}_{2}^{2}|\psi_{1}\rangle=0.
\end{equation}

\begin{figure}
\centering
\subfigure[\ mode 1]{
\begin{minipage}{4cm}
\centering
\includegraphics[width=4.2cm]{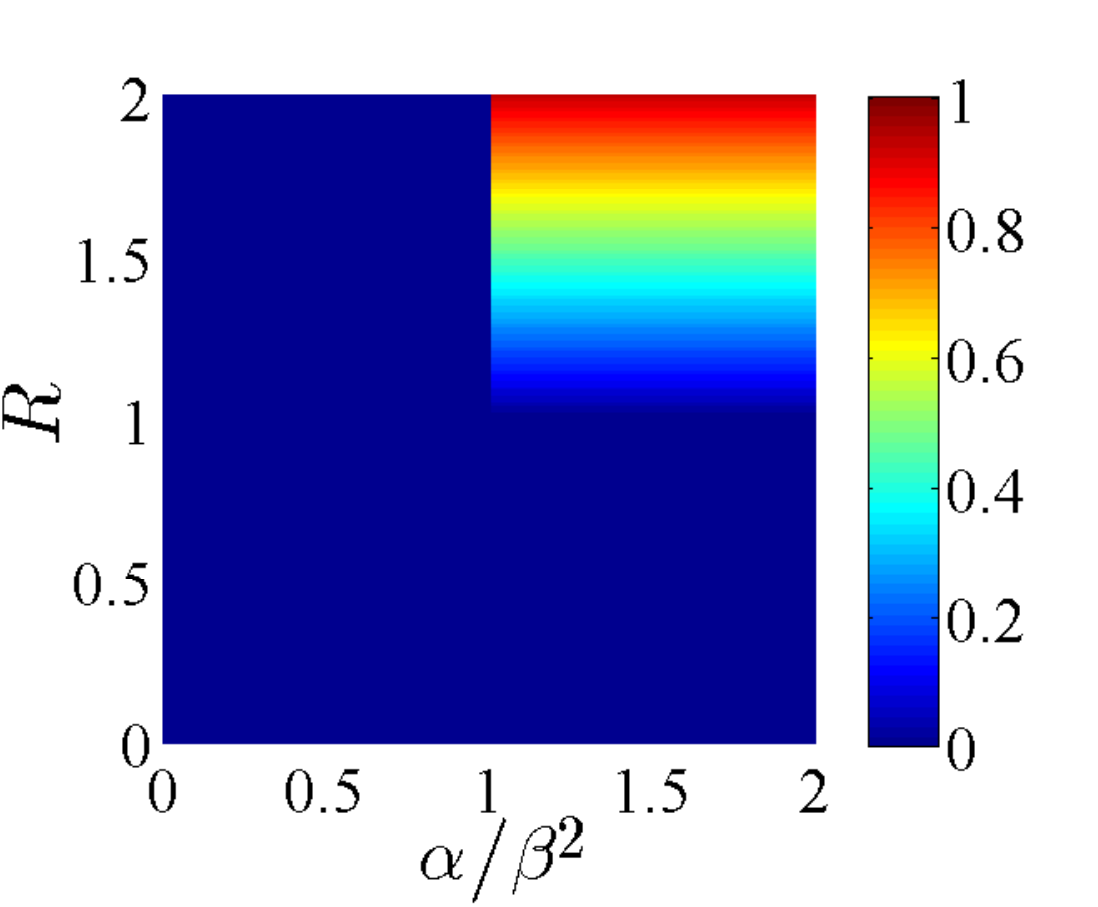}
\end{minipage}
}
\subfigure[\ mode 2]{
\begin{minipage}{4cm}
\centering
\includegraphics[width=4.2cm]{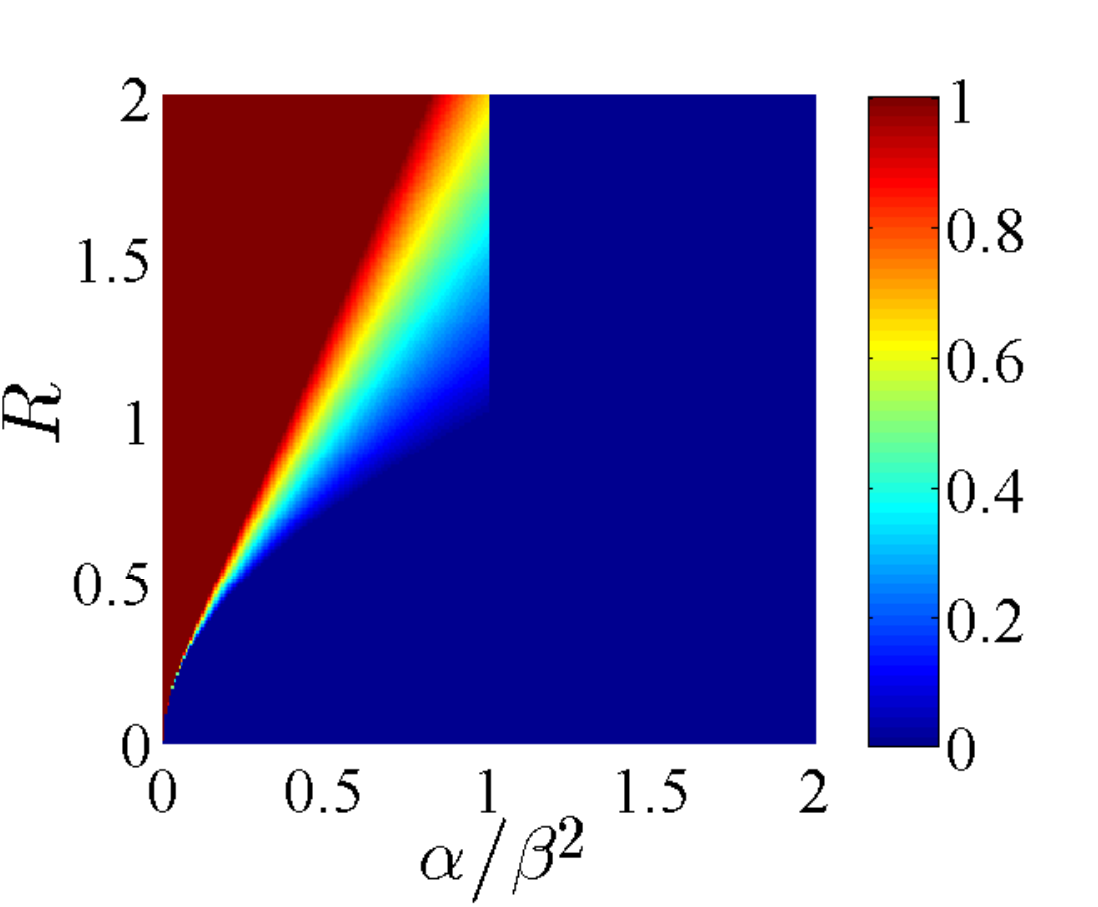}
\end{minipage}
}
\caption{Mean photon number as a function of $\alpha/\beta^{2}$ and $R$. In the normal phase below the critical points, the mean photon numbers of both modes are zero. Above the critical points, as for $\alpha/\beta^{2}<1$, the superradiant phase of $R>\sqrt{\alpha/\beta^{2}}$ is characterized by the mode-2 mean photon number of $\langle\hat{a}_{2}^{\dagger}\hat{a}_{2}\rangle/\eta$ (namely, $y_{2}$-type); As for $\alpha/\beta^{2}>1$, the superradiant phase of $R>1$ is characterized by the mode-1 mean photon number of $\langle\hat{a}_{1}^{\dagger}\hat{a}_{1}\rangle/\eta$ ($y_{1}$-type). ($\beta=1.2$).}
\label{phnum-analytical}
\end{figure}

We choose $\langle\hat{a}_{2}^{\dagger}\hat{a}_{2}\rangle/\eta$ and $\langle\hat{a}_{1}^{\dagger}\hat{a}_{1}\rangle/\eta$ as the order parameters for $\alpha<\beta^{2}$ and $\alpha>\beta^{2}$, respectively. Figure~\ref{phnum-analytical} presents the mean photon numbers of the two photon modes. One can see that the photon field is a vacuum in the normal phase and simultaneously the atom is at $|2\rangle$ for $\alpha<\beta^{2}$ and $|1\rangle$ for $\alpha>\beta^{2}$ inferring from Eqs.~\eqref{less-eigenstate} and \eqref{larger-eigenstate}. Above the critical points, both the photon field and the atom are excited and the so-called superradiant phase transition takes place. In each regime of $\alpha<\beta^{2}$ and $\alpha>\beta^{2}$, there is only one coupling involved. More precisely, only the coupling between the photon mode 2 and the transition $|2\rangle\leftrightarrow|3\rangle$ contributes to the ground state when $\alpha<\beta^{2}$, whereas only the coupling between the photon mode 1 and the transition $|1\rangle\leftrightarrow|3\rangle$ contributes to the ground state when $\alpha>\beta^{2}$. Actually, the two-mode three-level quantum Rabi model degenerates into the quantum Rabi model in these two parameter regimes. In each parameter regime, there exists a dark state, which is state $|1\rangle$ for $\alpha<\beta^{2}$ and $|2\rangle$ for $\alpha>\beta^{2}$, as seen in Eqs.~\eqref{less-eigenstate} and \eqref{larger-eigenstate}. The dark state was also studied in a semi-classical two-mode three-level quantum Rabi model \cite{Zhou2017}.

The critical exponent can be obtained by the deformation of the mean photon number with the reduced coupling strength $r=(R-R_{c})/R_{c}$,
\begin{equation}\label{larger-CriticalExponent}
\langle a_{1}^{\dagger}a_{1}\rangle/\eta=\frac{R^{4}-1}{4R^{2}} \propto r,
\end{equation}
\begin{equation}\label{less-CriticalExponent}
\langle a_{2}^{\dagger}a_{2}\rangle/\eta=\frac{\beta^{4}R^{4}-\alpha^{2}}{4\alpha^{2}\beta^{2}R^{2}} \propto r.
\end{equation}
They indicate that the critical exponent of the mean photon numbers is 1, which is the same as the quantum Rabi model \cite{Liu2017,Larson2017}. It suggests that the present model belongs to the same universality class with the quantum Rabi model.

\section{Scaling behavior}
The finite-size scaling behavior of continuous phase transitions is important near the critical point (the size here is characterized by the frequency ratio $\eta$). We derive the scaling function and critical exponents of this model. For finite $\eta$, after diagonalizing the atomic part, the Hamiltonian in Eq.~\eqref{Hpm} becomes ($\delta=0$)
\begin{eqnarray}\label{Hpm-}
\hat{H}'(\hat{y}_{1},\hat{y}_{2})=&&\frac{1}{2\eta^{2}}(\hat{p}_{y_{1}}^{2}+\alpha \hat{p}_{y_{2}}^{2})+\frac{1}{2}(\hat{y}_{1}^{2}+\alpha \hat{y}_{2}^{2}) \nonumber \\
&&+\frac{1}{2}[1-\sqrt{1+2R^{2}(\hat{y}_{1}^{2}+\beta^{2}\hat{y}_{2}^{2})}].
\end{eqnarray}

In the case of $\alpha<\beta^{2}$, the wavefunction near the critical point $R_{<,c}$ is very localized around $y_{2}=0$ when $\eta$ is very large. Thus, the Hamiltonian might be approximated by a second-order expansion in the vicinity of $y_{2}=0$,
\begin{equation}\label{less-Hpm}
\hat{H}'(\hat{y}_{2})\approx\frac{\alpha}{2\eta^{2}}\hat{p}_{y_{2}}^{2}-\alpha r\hat{y}_{2}^{2}+\frac{\alpha^{2}}{4}\hat{y}_{2}^{4}.
\end{equation}
Defining $\hat{y}_{2}=\eta^{-1/3}\hat{z}_{2}$ and $r=\eta^{-2/3}r'$, the Hamiltonian can be rewritten as
\begin{equation}
\hat{H}'(\hat{z}_{2})=\eta^{-4/3}(\frac{\alpha}{2}\hat{p}_{z_{2}}^{2}-\alpha r'\hat{z}_{2}^{2}+\frac{\alpha^{2}}{4}\hat{z}_{2}^{4}).
\end{equation}
In this respect there exists a rescaled ground-state wavefunction $\phi(\hat{z}_{2},r')$, which is independent of $\eta$. Based on this wavefunction, the mean photon number is
\begin{equation}
\langle\hat{a}_{2}^{\dagger}\hat{a}_{2}\rangle/\eta=\eta^{-2/3}\frac{1}{2}\langle\phi|\hat{z}_{2}^{2}|\phi\rangle.
\end{equation}

In the case of $\alpha>\beta^{2}$, the wavefunction near $R_{>,c}$ is also very localized around $y_{1}=0$ for a very large $\eta$. Likewise, we expand the Hamiltonian to the second order in the vicinity of $y_{1}=0$,
\begin{equation}\label{larger-Hpm}
\hat{H}'(\hat{y}_{1})\approx\frac{1}{2\eta^{2}}\hat{p}_{y_{1}}^{2}-r \hat{y}_{1}^{2}+\frac{1}{4}\hat{y}_{1}^{4}.
\end{equation}
Analogously defining $\hat{y}_{1}=\eta^{-1/3}\hat{z}_{1}$ and $r=\eta^{-2/3}r'$, we have
\begin{equation}
\hat{H}'(\hat{z}_{1})=\eta^{-4/3}(\frac{1}{2}\hat{p}_{z_{1}}^{2}-r' \hat{z}_{1}^{2}+\frac{1}{4}\hat{z}_{1}^{4}).
\end{equation}
In this case, the mean photon number is
\begin{equation}
\langle\hat{a}_{1}^{\dagger}\hat{a}_{1}\rangle/\eta=\eta^{-2/3}\frac{1}{2}\langle\phi|\hat{z}_{1}^{2}|\phi\rangle.
\end{equation}
The detailed derivation of the Hamiltonian from Eq.~\eqref{Hpm-} to Eqs.~\eqref{less-Hpm} and \eqref{larger-Hpm} is given in Appendix B.

Based on the above analytical derivation, the scaling function for the mean photon number of the two-mode three-level quantum Rabi model is obtained as follows,
\begin{equation}\label{OurScalingFunction}
N(\eta,r)=\eta^{-2/3}f(\eta^{2/3}r).
\end{equation}
where $N(\eta,r)$ is the photon numbers of $\langle\hat{a}_{1}^{\dagger}\hat{a}_{1}\rangle/\eta$ and $\langle\hat{a}_{2}^{\dagger}\hat{a}_{2}\rangle/\eta$.

According to the standard finite-size scaling law \cite{Privman1984}, the divergent correlation length at critical points enables the scaling behavior of a physical quantity $P$ at different finite $\eta$:
\begin{equation}\label{StandardScalingFunction}
P(\eta,r)=\eta^{-\kappa/\nu}f(\eta^{1/\nu}r),
\end{equation}
where $\kappa$ is the critical exponent of $P$ ($P\propto r^{\kappa}$), $\nu$ is the critical exponent of the correlation length ($\xi\propto r^{-\nu}$). Comparing Eq.~\eqref{OurScalingFunction} with Eq.~\eqref{StandardScalingFunction}, we infer that $\kappa$ and $\nu$ of the mean photon number for the two-mode three-level quantum Rabi model are 1 and 3/2, respectively. The finite-size critical scaling exponent $\kappa$ of the mean photon numbers is the same as that obtained from the limit $\eta\rightarrow\infty$, as is concluded from Eqs.~\eqref{larger-CriticalExponent} and \eqref{less-CriticalExponent}.

To check our analytical prediction, we numerically diagonalize the two-mode three-level quantum Rabi model. We calculate the mean photon numbers of the two photon modes in the case of $\delta=0$. To determine critical points and critical exponents numerically, we take the logarithm of Eq.~\eqref{StandardScalingFunction} as
\begin{equation}
\ln P(\eta,r)=(-\kappa/\nu)\ln\eta+\ln f(\eta^{1/\nu}r).
\end{equation}
At the critical point $r=0$, $\ln P(\eta,r)$ and $\ln\eta$ are linearly related as
\begin{equation}
\ln P(\eta,0)=(-\kappa/\nu)\ln\eta+\ln f(0).
\end{equation}
The scaling parameter $\kappa/\nu$ can be determined by a linear fitting. The remaining parameter $\nu$ can be obtained by a collapse of the data points with different $\eta$ values onto a single scaled curve.

\begin{figure}
\centering
\subfigure[\ linear fitting]{
\begin{minipage}{4cm}
\centering
\includegraphics[width=4.2cm]{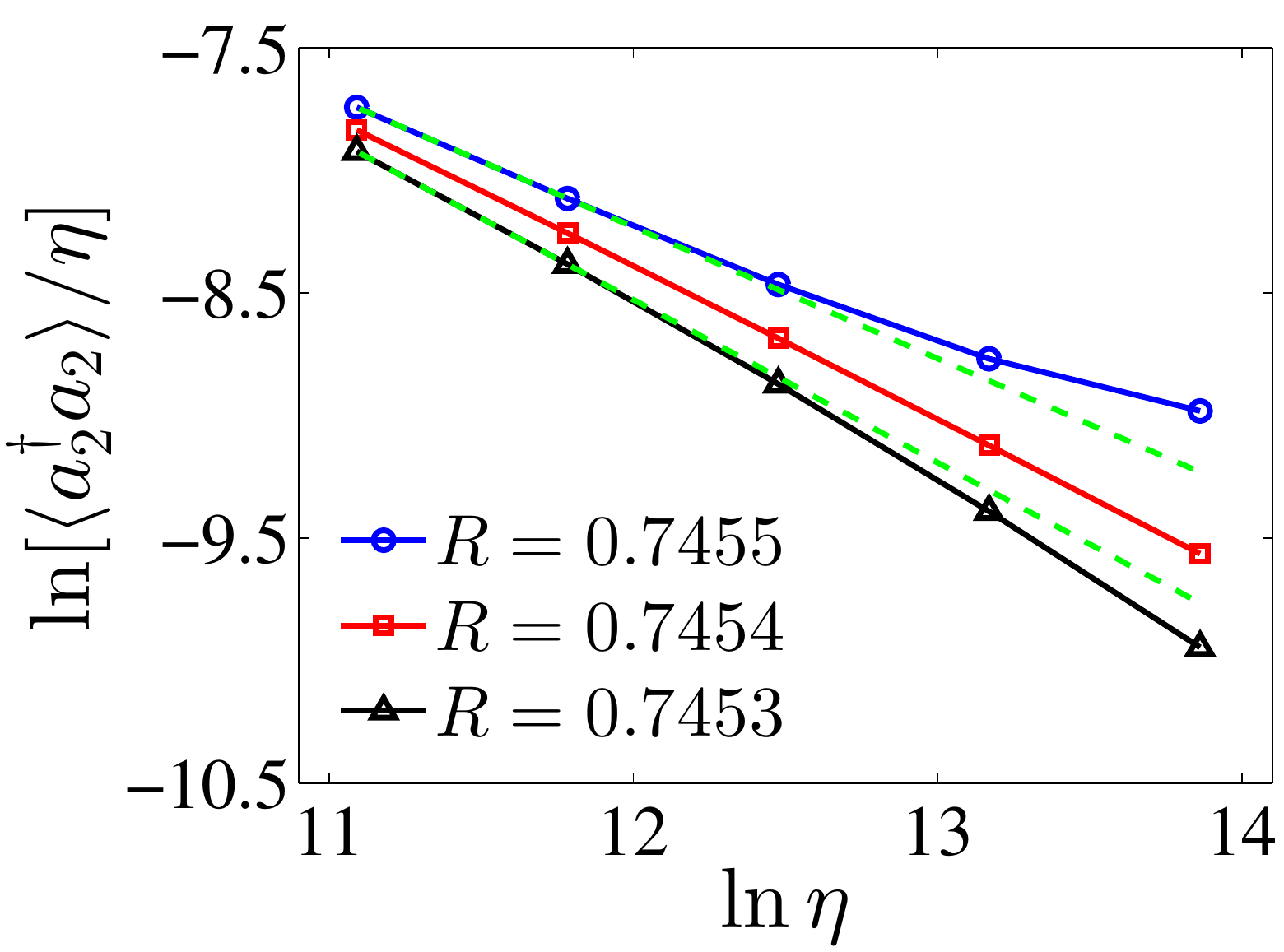}
\end{minipage}
}
\subfigure[\ scaling invariance]{
\begin{minipage}{4cm}
\centering
\includegraphics[width=4.2cm]{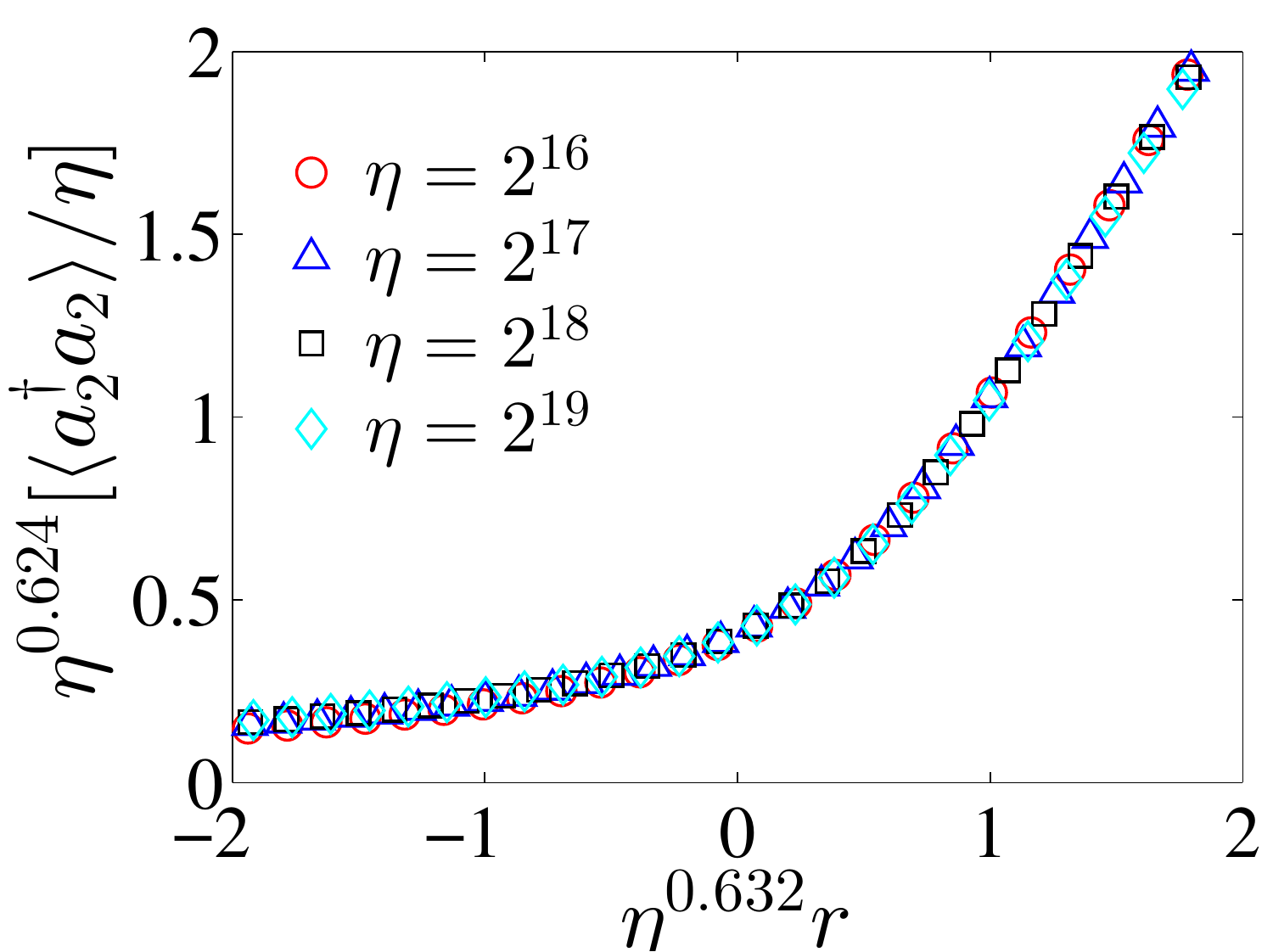}
\end{minipage}
}
\caption{$\alpha<\beta^{2}$ (a) $\ln[\langle\hat{a}_{2}^{\dagger}\hat{a}_{2}\rangle/\eta]$ as a function of $\ln\eta$. The red-square line exhibits the linear relation, which indicates the critical point of 0.7454. Two green dashed straight lines assist us to observe. (b) Scaling of $\langle\hat{a}_{2}^{\dagger}\hat{a}_{2}\rangle/\eta$ makes all the data points of different $\eta$ values collapse onto a single curve. ($\alpha=0.8$ and $\beta=1.2$).}
\label{less-scaling}
\end{figure}

\begin{figure}
\centering
\subfigure[\ linear fitting]{
\begin{minipage}{4cm}
\centering
\includegraphics[width=4.2cm]{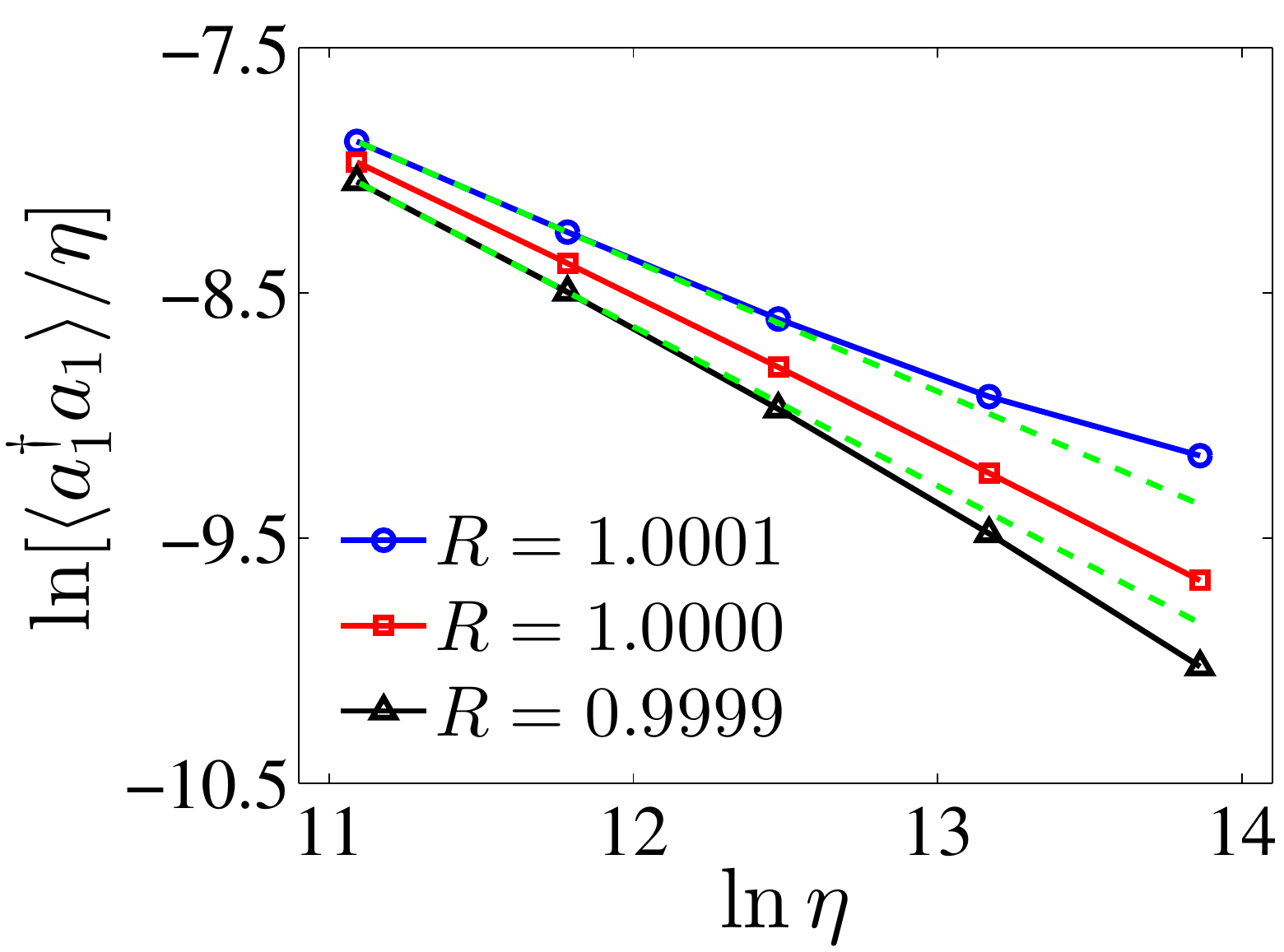}
\end{minipage}
}
\subfigure[\ scaling invariance]{
\begin{minipage}{4cm}
\centering
\includegraphics[width=4.2cm]{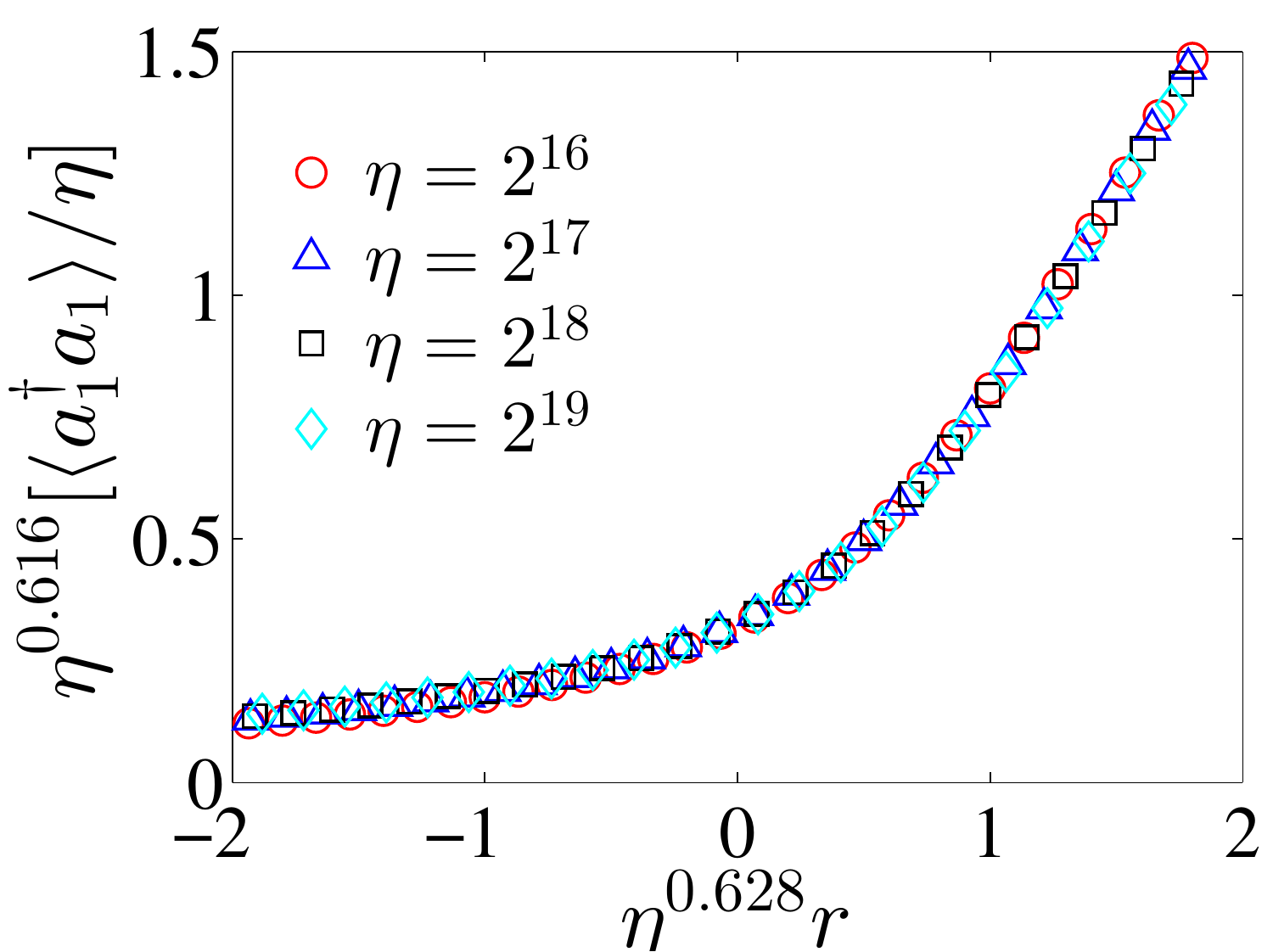}
\end{minipage}
}
\caption{$\alpha>\beta^{2}$ (a) $\ln[\langle\hat{a}_{1}^{\dagger}\hat{a}_{1}\rangle/\eta]$ as a function of $\ln\eta$. The red-square line shows the linear relation at the critical point of 1.0000. (b) Scaling of $\langle\hat{a}_{1}^{\dagger}\hat{a}_{1}\rangle/\eta$ makes all the data points of different $\eta$ values collapse onto a single curve. ($\alpha=1.2$ and $\beta=0.8$).}
\label{larger-scaling}
\end{figure}

Figures~\ref{less-scaling}(a) and \ref{larger-scaling}(a) show the logarithm of the mean photon number as a function of the logarithm of $\eta$ for the two regimes of $\alpha<\beta^{2}$ and $\alpha>\beta^{2}$, respectively. The red-square line represents the linear behavior and the determined critical points are $R_{<,c}=0.7454$ and $R_{>,c}=1.0000$, which confirm the analytical critical points of $R_{<,c}=\sqrt{\alpha/\beta^{2}}$ and $R_{>,c}=1$, respectively. The fitted values of $\kappa/\nu$ are 0.624 and 0.616, and $\nu$ is correspondingly obtained to be 1.582 and 1.592, which reproduce the analytical values $\kappa/\nu=2/3$ and $\nu=3/2$. One can see from Figs.~\ref{less-scaling}(b) and \ref{larger-scaling}(b) that the mean photon numbers for different $\eta$ collapse onto a well-defined single curve. It indicates that the analytical solution reveals the correct ground-state phase diagram of the two-mode three-level quantum Rabi model and captures its scaling invariance near the critical points.

\section{Conclusions}
Based on the analytical solution and the numerical diagonalization, we attain the ground-state phase diagram, scaling function, and critical exponents of the two-mode three-level quantum Rabi model. The phase diagram is divided into the three regions of one normal phase and two superradiant phases. The phase transitions could take place by adjusting the frequency ratio of the two photon modes and the relative strength of the two photon-atom couplings. The normal-superradiant quantum phase transitions are found to be second order and related to the spontaneous breaking of $Z_{2}$ symmetry. In addition, the model undergoes a first-order phase transition across the boundary line between the two superradiant phases, where a spontaneous continuous $U(1)$ symmetry breaking is discovered. Different from the traditional phase transitions in the thermodynamic limit, here the quantum phase transition is realized alternatively when the frequency ratio $\eta$ of the atomic transition and the photon field approaches infinity. The finite-$\eta$ scaling function is derived and the obtained critical exponent of the mean photon number is the same as that determined in the limit $\eta\rightarrow\infty$. Based on this, the universality class of this model is identified. This work is helpful in further understanding the quantum phase transitions and exploring some potential applications for such single-atom systems. The present results are obtained from the degenerate case for two lowest states, the general case of the two-mode three-level quantum Rabi model needs further exploration.

\begin{acknowledgments}
This work was supported by NSFC (Grants No. 11604009, No. 11705008, No. 11474211, No. 11674139, No. 11834005, and No. 11504298) and NSAF (Grant No. U1530401). D.X. was also supported by the Beijing Institute of Technology Research Fund Program for Young Scholars.
\end{acknowledgments}

\appendix

\section{Derivation of the ground-state energy}
To determine the ground-state energy of Eq.~\eqref{Eg}, we calculate its first derivative with respect to $y_{1}$ and $y_{2}$, and let both them equal zero
\begin{equation}
\begin{cases}
\frac{\partial E({y}_{1},{y}_{2})}{\partial y_{1}}=0 \\
\frac{\partial E({y}_{1},{y}_{2})}{\partial y_{2}}=0.
\end{cases}
\end{equation}
It yields
\begin{equation}
\begin{cases}
y_{1}(1-\frac{R^{2}}{\sqrt{1+2R^{2}(y_{1}^{2}+\beta^{2}y_{2}^{2})}})=0 \\
y_{2}(\alpha-\frac{R^{2}\beta^{2}}{\sqrt{1+2R^{2}(y_{1}^{2}+\beta^{2}y_{2}^{2})}})=0.
\end{cases}
\end{equation}
The solutions have four cases:

\noindent Case 1:
\begin{equation}
\begin{cases}
y_{1}=0 \\
y_{2}=0;
\end{cases}
\end{equation}
the corresponding ground-state energy $E=0$.

\noindent Case 2:
\begin{equation}
\begin{cases}
y_{1}=0 \\
\alpha-\frac{R^{2}\beta^{2}}{\sqrt{1+2R^{2}(y_{1}^{2}+\beta^{2}y_{2}^{2})}}=0;
\end{cases}
\end{equation}
we thus have
\begin{equation}
y_{2,\pm}=\pm\sqrt{\frac{\beta^{4}R^{4}-\alpha^{2}}{2\alpha^{2}\beta^{2}R^{2}}}.
\end{equation}
This solution exists only when $R\geqslant\sqrt{\frac{\alpha}{\beta^{2}}}$, and the corresponding energy $E=-\frac{1}{4}(\frac{\alpha}{\beta^{2}R^{2}}+\frac{\beta^{2}R^{2}}{\alpha})+\frac{1}{2}\leqslant0$.

\noindent Case 3:
\begin{equation}
\begin{cases}
1-\frac{R^{2}}{\sqrt{1+2R^{2}(y_{1}^{2}+\beta^{2}y_{2}^{2})}}=0 \\
y_{2}=0;
\end{cases}
\end{equation}
hence,
\begin{equation}
y_{1,\pm}=\pm\sqrt{\frac{R^{4}-1}{2R^{2}}}.
\end{equation}
It exists only when $R\geqslant1$, and the corresponding energy $E=-\frac{1}{4}(\frac{1}{R^{2}}+R^{2})+\frac{1}{2}\leqslant0$.

\noindent Case 4:
\begin{equation}
\begin{cases}
1-\frac{R^{2}}{\sqrt{1+2R^{2}(y_{1}^{2}+\beta^{2}y_{2}^{2})}}=0 \\
\alpha-\frac{R^{2}\beta^{2}}{\sqrt{1+2R^{2}(y_{1}^{2}+\beta^{2}y_{2}^{2})}}=0
\end{cases}
\end{equation}
requires $\alpha=\beta^{2}$. On substituting this relation into Eq.~\eqref{Eg}, we have
\begin{equation}
E({y}_{1},{y}_{2})=\frac{1}{2}(y_{1}^{2}+\beta^{2}y_{2}^{2})+\frac{1}{2}[1-\sqrt{1+2R^{2}(y_{1}^{2}+\beta^{2}y_{2}^{2})}].
\end{equation}
Defining $y_{1}^{2}+\beta^{2}y_{2}^{2}=S$, the ground-state energy expression becomes
\begin{equation}
E(S)=\frac{1}{2}S+\frac{1}{2}[1-\sqrt{1+2R^{2}S}].
\end{equation}
We let its first-order derivative $\frac{{\rm d}E(S)}{{\rm d}S}=0$, then
\begin{equation}
S=\frac{R^{4}-1}{2R^{2}}
\end{equation}
It indicates a continuous set of $y_{1}$ and $y_{2}$ with the degenerate energy of $E=-\frac{1}{4}(\frac{1}{R^{2}}+R^{2})+\frac{1}{2}\leqslant0$, which exists also when $R\geqslant1$

\begin{widetext}

To summarize, we get the ground-state energies as follows,
\begin{table}[!h]
  \centering
  \begin{tabular}{|c|lll|}
  \hline
  \multirow{2}{*}{$\alpha<\beta^{2}$} & $R\leqslant\sqrt{\frac{\alpha}{\beta^{2}}}$ & $\{y_{1}=0,y_{2}=0\}$ & $E_{0}=0$ \\
  & $R>\sqrt{\frac{\alpha}{\beta^{2}}}$ & $\{y_{1}=0,y_{2,\pm}=\pm\sqrt{\frac{\beta^{4}R^{4}-\alpha^{2}}{2\alpha^{2}\beta^{2}R^{2}}}\}$ & $E_{0}=-\frac{1}{4}(\frac{\alpha}{\beta^{2}R^{2}}+\frac{\beta^{2}R^{2}}{\alpha})+\frac{1}{2}$ \\
  \hline
  \multirow{2}{*}{$\alpha>\beta^{2}$} & $R\leqslant1$ & $\{y_{1}=0,y_{2}=0\}$ & $E_{0}=0$ \\
  & $R>1$ & $\{y_{1,\pm}=\pm\sqrt{\frac{R^{4}-1}{2R^{2}}},y_{2}=0\}$ & $E_{0}=-\frac{1}{4}(\frac{1}{R^{2}}+R^{2})+\frac{1}{2}$ \\
  \hline
  \multirow{2}{*}{$\alpha=\beta^{2}$} & $R\leqslant1$ & $\{y_{1}=0,y_{2}=0\}$ & $E_{0}=0$ \\
  & $R>1$ & $\{y_{1}^{2}+\beta^{2}y_{2}^{2}=\frac{R^{4}-1}{2R^{2}}\}$ & $E_{0}=-\frac{1}{4}(\frac{1}{R^{2}}+R^{2})+\frac{1}{2}$ \\
  \hline
  \end{tabular}
\end{table}

\noindent They are verified by the positive second-order derivatives of $\frac{\partial^{2} E({y}_{1},{y}_{2})}{\partial y_{1}^{2}}>0$ and $\frac{\partial^{2} E({y}_{1},{y}_{2})}{\partial y_{2}^{2}}>0$ at the corresponding $y_{1}$ and $y_{2}$.

\end{widetext}

\section{Derivation of the finite-$\eta$ Hamiltonian}
For a large enough but finite $\eta$, we derive the Hamiltonian of this model near the critical point from Eq.~\eqref{Hpm-}.

When $\alpha<\beta^{2}$, the superradiant phase is $y_{2}$-type in view of the ground-state phase diagram (see Fig.~\ref{PhaseDiagram}). Although the contributions of the $\hat{y}_{1}$ terms are not strictly zero for finite $\eta$, ignoring the $\hat{y}_{1}$ terms in Eq.~\eqref{Hpm-} should be a good approximation when $\eta$ is large enough. The Hamiltonian is given by
\begin{equation}
\hat{H}'(\hat{y}_{2})=\frac{\alpha}{2\eta^{2}}\hat{p}_{y_{2}}^{2}+\frac{\alpha}{2}\hat{y}_{2}^{2}+\frac{1}{2}[1-\sqrt{1+2\beta^{2}R^{2}\hat{y}_{2}^{2}}].
\end{equation}
The wavefunction of this Hamiltonian is very localized because the contributions of the momentum term becomes very small when $\eta$ is very large. In particular, the wavefunction is localized around $y_{2}=0$ near the critical point. Thus, we express the Hamiltonian by a second-order expansion in the vicinity of $y_{2}=0$,
\begin{equation}
\hat{H}'(\hat{y}_{2})\approx\frac{\alpha}{2\eta^{2}}\hat{p}_{y_{2}}^{2}+\frac{1}{2}(\alpha-\beta^{2}R^{2})\hat{y}_{2}^{2}+\frac{1}{4}\beta^{4}R^{4}\hat{y}_{2}^{4}.
\end{equation}
Considering that the coupling strength $R$ is close to the critical point, namely, $R\approx\sqrt{\alpha/\beta^{2}}$, the Hamiltonian can be further processed as
\begin{equation}
\hat{H}'(\hat{y}_{2})\approx\frac{\alpha}{2\eta^{2}}\hat{p}_{y_{2}}^{2}+\sqrt{\alpha}(\sqrt{\alpha}-\beta R)\hat{y}_{2}^{2}+\frac{\alpha^{2}}{4}\hat{y}_{2}^{4}.
\end{equation}
Recalling the reduced coupling strength $r=(R-R_{c})/R_{c}$, we have
\begin{equation}
\hat{H}'(\hat{y}_{2})\approx\frac{\alpha}{2\eta^{2}}\hat{p}_{y_{2}}^{2}-\alpha r\hat{y}_{2}^{2}+\frac{\alpha^{2}}{4}\hat{y}_{2}^{4}.
\end{equation}

When $\alpha>\beta^{2}$, the superradiant phase is $y_{1}$-type. Similarly, when $\eta$ is large enough, the Hamiltonian becomes
\begin{equation}
\hat{H}'(\hat{y}_{1})=\frac{1}{2\eta^{2}}\hat{p}_{y_{1}}^{2}+\frac{1}{2}\hat{y}_{1}^{2}+\frac{1}{2}[1-\sqrt{1+2R^{2}\hat{y}_{1}^{2}}].
\end{equation}
Its wavefunction near the critical point is very localized around $y_{1}=0$ when $\eta$ is very large, and we expand the Hamiltonian to the second order in the vicinity of $y_{1}=0$,
\begin{equation}
\hat{H}'(\hat{y}_{1})\approx\frac{1}{2\eta^{2}}\hat{p}_{y_{1}}^{2}+\frac{1}{2}(1-R^{2})\hat{y}_{1}^{2}+\frac{1}{4}R^{4}\hat{y}_{1}^{4}.
\end{equation}
Using $R\approx1$, we obtain
\begin{equation}
\hat{H}'(\hat{y}_{1})\approx\frac{1}{2\eta^{2}}\hat{p}_{y_{1}}^{2}+(1-R)\hat{y}_{1}^{2}+\frac{1}{4}\hat{y}_{1}^{4}.
\end{equation}
Inserting the reduced coupling strength $r=R-1$ into the above equation, we have
\begin{equation}
\hat{H}'(\hat{y}_{1})\approx\frac{1}{2\eta^{2}}\hat{p}_{y_{1}}^{2}-r\hat{y}_{1}^{2}+\frac{1}{4}\hat{y}_{1}^{4}.
\end{equation}

\end{document}